\documentclass{aastex61}

\received{}
\revised{}
\accepted{}
\submitjournal{ApJ}

\begin{document}

\title{$^{13}$C isotopic fractionation of HC$_{3}$N in two starless cores: L1521B and L134N (L183)}

\correspondingauthor{Kotomi Taniguchi}
\email{kotomi.taniguchi@nao.ac.jp}

\author{Kotomi Taniguchi}
\altaffiliation{Research Fellow of Japan Society for the Promotion of Science}
\affiliation{Department of Astronomical Science, School of Physical Science, SOKENDAI (The Graduate University for Advanced Studies), Osawa, Mitaka, Tokyo 181-8588, Japan}
\affiliation{Nobeyama Radio Observatory, National Astronomical Observatory of Japan, Minamimaki, Minamisaku, Nagano 384-1305, Japan}

\author{Hiroyuki Ozeki}
\affiliation{Department of Environmental Science, Faculty of Science, Toho University, Miyama, Funabashi, Chiba 274-8510, Japan}

\author{Masao Saito}
\affiliation{National Astronomical Observatory of Japan, Osawa, Mitaka, Tokyo 181-8588, Japan}
\affiliation{Department of Astronomical Science, School of Physical Science, SOKENDAI (The Graduate University for Advanced Studies), Osawa, Mitaka, Tokyo 181-8588, Japan}



\begin{abstract}

We observed the $J=5-4$ rotational lines of the normal species and three $^{13}$C isotopologues of HC$_{3}$N at the 45 GHz band toward two low-mass starless cores, L1521B and L134N (L183), using the Nobeyama 45 m radio telescope in order to study the main formation pathways of HC$_{3}$N in each core.
The abundance ratios of the three $^{13}$C isotopologues in L1521B are derived to be [H$^{13}$CCCN]:[HC$^{13}$CCN]:[HCC$^{13}$CN] = 0.98 ($\pm 0.14$) : 1.00 : 1.52 ($\pm 0.16$) ($1 \sigma$).
The fractionation pattern is consistent with that at the cyanopolyyne peak in Taurus Molecular Cloud-1.
This fractionation pattern suggests that the main formation pathway of HC$_{3}$N is the neutral-neutral reaction between C$_{2}$H$_{2}$ and CN. 
On the other hand, their abundance ratios in L134N are found to be [H$^{13}$CCCN]:[HC$^{13}$CCN]:[HCC$^{13}$CN]= 1.5 ($\pm 0.2$) : 1.0 : 2.1 ($\pm 0.4$) ($1 \sigma$), which are different from those in L1521B.
From this fractionation pattern, we propose that the reaction between HNC and CCH is a possible main formation pathway of HC$_{3}$N in L134N.
We find out that the main formation pathways of the same molecule are not common even in the similar physical conditions.
We discuss the possible factors to make a difference in fractionation pattern between L134N and L1521B/TMC-1.
\end{abstract}

\keywords{astrochemistry --- ISM: individual objects (L1521B, L134N (L183)) --- ISM: molecules}



\section{Introduction} \label{sec:intro}

Carbon-chain molecules have been detected usually in young molecular clouds.
They account for $\sim 40$\% of the approximately 200 molecules detected in the interstellar medium and circumstellar shells.
They are formed from C or C$^{+}$ in the early stages of molecular clouds \citep{1992ApJ...392...551S}, and decrease as clouds evolve by depletion onto dust grains or reactions with ions such as H$^{+}$, He$^{+}$, or oxygen atoms \citep{2013ChRv...113...8981S}.
However, their formation pathways of each individual molecule are still controversial due to lacks of laboratory experiments, because carbon-chain molecules and their ions are unstable. 

One method to study the main formation pathways of carbon-chain molecules is deriving their $^{13}$C isotopic fractionation by radio astronomical observations.
Several observations to derive the $^{13}$C isotopic fractionation of carbon-chain molecules have been carried out mainly at the cyanopolyyne peak in Taurus Molecular Cloud-1 (TMC-1 CP; $d = 140$ pc) in order to investigate their main formation pathways (HC$_{3}$N \citep{1998A&A...329.1156T}, HC$_{5}$N \citep{2016ApJ...817...147T}, CCS \citep{2007ApJ...663...1174S}, CCH \citep{2010ApJ...512...A31S}, C$_{3}$S and C$_{4}$H \citep{2013JPCA...117...9831S}).
The abundance ratios of HC$_{3}$N in TMC-1 CP were derived to be [H$^{13}$CCCN]:[HC$^{13}$CCN]:[HCC$^{13}$CN] = 1.0 : 1.0 : 1.4 ($\pm 0.2$) ($1 \sigma$) \citep{1998A&A...329.1156T}.
The authors suggested that the main formation pathway of HC$_{3}$N in TMC-1 CP is the neutral-neutral reaction between C$_{2}$H$_{2}$ and CN, based on the abundance ratios.
On the other hand, \citet{2016ApJ...817...147T} proposed that the main formation mechanism of HC$_{5}$N in TMC-1 CP is the ion-molecule reactions between hydrocarbon ions (C$_{5}$H$_{m}^{+}$) and nitrogen atoms followed by the electron recombination reactions, because significant differences in abundance among the five $^{13}$C isotopologues of HC$_{5}$N were not recognized.
Recently, \citet{2017arXiv170608662T} further confirmed the main formation mechanism of HC$_{5}$N using the $^{14}$N/$^{15}$N ratio of HC$_{5}$N.

The main formation pathways of HC$_{3}$N were further investigated in the low-mass star-forming region L1527 ($d=140$ pc) and the high-mass star-forming region G28.28-0.36 ($d=3$ kpc) using the Nobeyama 45 m radio telescope \citep{2016ApJ...830...106T}.
L1527 is one of the warm carbon chain chemistry (WCCC) sources where carbon-chain molecules are formed from CH$_{4}$ evaporated from grain mantles in the warm gas, $T \simeq 20-30$ K \citep[e.g.][]{2013ChRv...113...8981S}.
G28.28-0.36 is a high-mass star-forming core associated with the 6.7 GHz methanol maser.
\citet{2017ApJ...844...68T} found that HC$_{5}$N exists in the warm gas around the massive young stellar object, and suggested efficient formation mechanisms of carbon chains in the high-mass star-forming region from the high HC$_{5}$N abundance.
The $^{13}$C isotopic fractionation patterns in the both star-forming regions are the same one as TMC-1 CP; the abundances of H$^{13}$CCCN and HC$^{13}$CCN are comparable with each other, and the abundance of HCC$^{13}$CN is higher than the other two isotopologues.
From these results, \citet{2016ApJ...830...106T} proposed that the main formation pathways in L1527 and G28.28-0.36 are the neutral-neutral reaction between C$_{2}$H$_{2}$ and CN, which is the same one in TMC-1 CP.
They also suggested that the primary formation pathway of HC$_{3}$N may be common in from low-mass prestellar to high-mass star-forming cores.
%\citet{2016ApJ...833..291A} carried out the same observations only toward L1527 obtaining spectra with higher signal-to-noise ratios, and reached the same fractionation pattern.

There is a possibility that TMC-1 CP shows unusual starless core chemistry, because long carbon-chain molecules are extraordinarily abundant in TMC-1 CP \citep[e.g.][]{2004PASJ...56...69K}.
\citet{2000ApJ...535..256M} suggested that an IRAS source located at the northern part of TMC-1 may affect the chemistry in TMC-1 cloud.
In order to confirm whether the neutral-neutral reaction of C$_{2}$H$_{2}$ + CN is the common main formation pathway of HC$_{3}$N in low-mass prestellar cores, we need to investigate in other starless cores which are not associated with any IRAS sources.

In the present paper, we report the observations of the normal species and the three $^{13}$C isotopologues of HC$_{3}$N toward two starless cores, L1521B and L134N, with the Nobeyama 45 m radio telescope.
The main purpose is to study the main formation pathways of HC$_{3}$N in different chemical conditions with the similar physical conditions.
L1521B is a low-mass cold starless core at the very early stage of physical and chemical evolution and is known to be rich in carbon-chain molecules like TMC-1 CP \citep{2004ApJ...617..399H}.
L134N, which is also known as L183, is one of the well-studied cold starless cores \citep[e.g.][]{2000ApJ...542..870D, 2003A&A...406L..59P, 2004A&A...417..605P, 2005A&A...429..181P}.
Although the temperature and density in L134N are similar to those in TMC-1 CP ($T_{\rm kin} \sim 10$ K, $n \sim 10^{4} - 10^{5}$ cm$^{-3}$), the chemical composition is different from that in TMC-1 CP.
Abundances of carbon-chain molecules in L134N are lower than those in TMC-1 CP, whereas L134N is richer in oxygen-rich molecules and NH$_{3}$ \citep{2000ApJ...542..870D}.
It is considered that L134N is chemically more evolved than TMC-1 CP \citep{2000ApJ...542..870D}.
Based on the observational results of the $^{13}$C isotopic fractionation, we discuss the main formation pathways in each core in Section \ref{sec:d1}.
In Section \ref{sec:d3}, we discuss possible factors causing the differences in fractionation pattern between L134N and L1521B/TMC-1 CP.
\\
\\
\section{Observations} \label{sec:obs}
\subsection{Observations toward L1521B}

The observations toward L1521B were carried out in 2016 December with the Nobeyama 45 m radio telescope.
The $J=5-4$ rotational lines of the normal species and the three $^{13}$C isotopologues of HC$_{3}$N at the 45 GHz band were observed simultaneously.
We employed the position-switching mode, and the scan pattern was 20 and 20 seconds for on-source and off-source positions, respectively.
The observed position was ($\alpha_{2000}$, $\delta_{2000}$) = (04$^{\rm h}$24$^{\rm m}$12\fs67, +26\arcdeg36\arcmin52\farcs8).
The off-source position was set to be ($\alpha$, $\delta$) = (+4\arcmin, +4\arcmin) away from the observed position.

We used the Z45 receiver, which is a dual-polarization HEMT amplifier receiver \citep{2015PASJ...67..117N}.
The beam size (HPBW) and the main beam efficiency ($\eta_{\rm {B}}$) of the Z45 receiver were 37$^{\prime \prime}$ and 71\%, respectively. 
The system temperatures were from 115 to 140 K depending on the weather conditions and elevations.
We used the SAM45 FX-type digital correlator \citep{2012PASJ...64...29K} in frequency setting whose bandwidth and frequency resolution are 125 MHz and 30.52 kHz, respectively.
The frequency resolution corresponds to the velocity resolution of 0.2 km s$^{-1}$ at 45 GHz.

We checked telescope pointing by observing SiO maser line ($J=1-0$) from NML Tau at ($\alpha_{2000}$, $\delta_{2000}$) = (03$^{\rm h}$53$^{\rm m}$28\fs86, +11\arcdeg24\arcmin22\farcs4) using the Z45 receiver every 2 hr.
The pointing  errors were less than 3\arcsec.

\subsection{Observations toward L134N}

We carried out observations toward L134N in 2017 February, April, and May\footnote{The observations in April and May were carried out without the master collimator driving system. The problem does not affect the results of the $^{13}$C isotopic fractionation and the $^{12}$C/$^{13}$C ratios, because all of the target lines including the normal species were observed simultaneously and the pointing errors were offset.}.
The target lines were the same ones as the observations toward L1521B.
The observed position was ($\alpha_{2000}$, $\delta_{2000}$) = (15$^{\rm h}$54$^{\rm m}$12\fs72, -02\arcdeg49\arcmin47\farcs4).
The off-source position was set to be +3$^{\prime}$ away in the right ascension.
We employed the smoothed bandpass calibration (SBC) method \citep{2012PASJ...64..118Y} to reduce observing time for off-source position.
We set the scan pattern as 20 and 5 seconds for on-source and off-source positions, respectively.
We applied 150-channel smoothing for off-source spectra.

We used the Z45 receiver and the SAM45 FX-type digital correlator in frequency setting whose bandwidth and frequency resolution are 31.25 MHz and 7.63 kHz, respectively. 
We conducted smoothing in the velocity direction of 0.2 km s$^{-1}$ in the final spectra.
The system temperatures were between 115 and 220 K, depending on the weather conditions and elevations.

The telescope pointing was checked by observing the SiO maser line ($J=1-0$) from R-Ser at ($\alpha_{2000}$, $\delta_{2000}$) = (15$^{\rm h}$50$^{\rm m}$41\fs735, +15\arcdeg08\arcmin01\farcs42) using the Z45 receiver every 1 hr.
The pointing errors in February were approximately within 3\arcsec.
The pointing errors in April and May were estimated at $\sim 10^{\prime \prime}$ due to the absence of the master collimator driving system.

\section{Results and Analysis}
\subsection{Results}

\begin{figure}
\figurenum{1}
\plotone{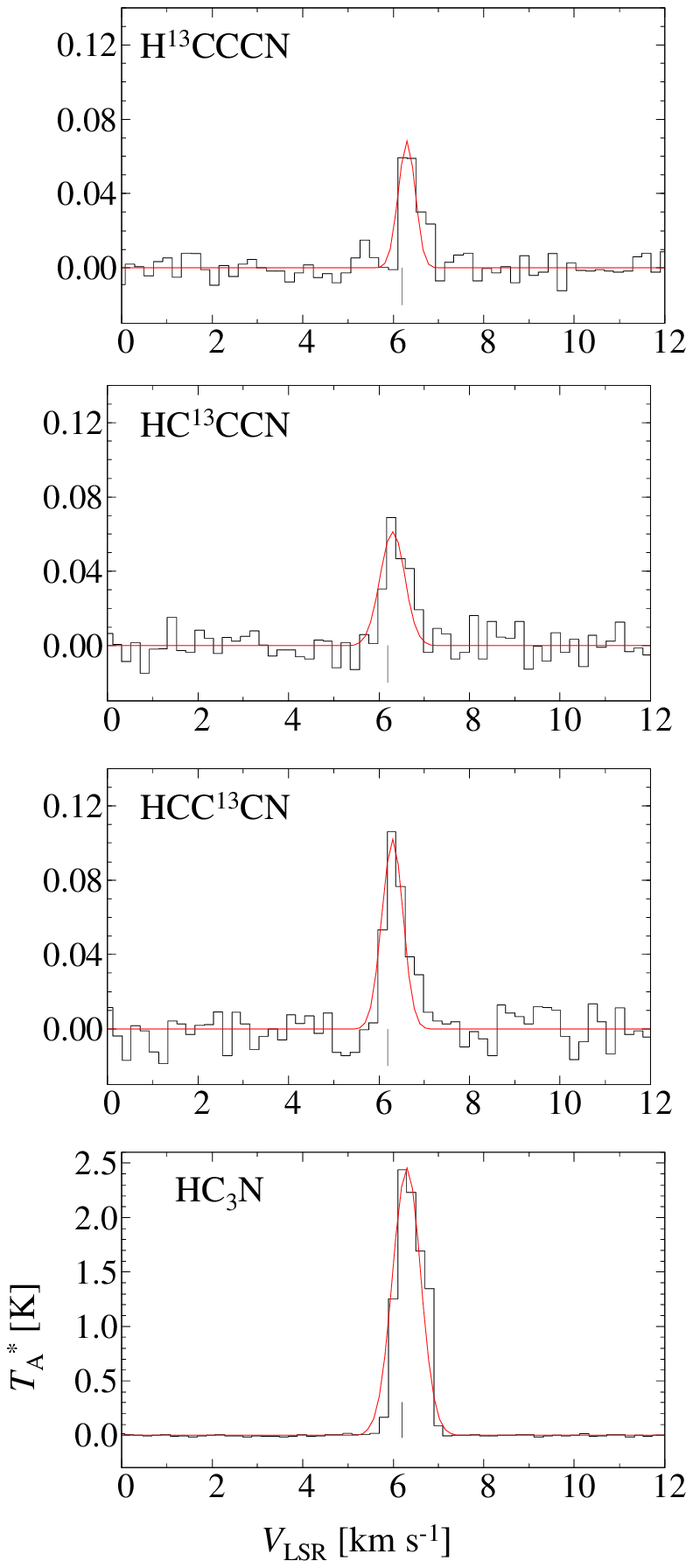}
\caption{Spectra of the normal species and the three $^{13}$C isotopologues of HC$_{3}$N in L1521B. The vertical lines show the systemic velocity ($V_{\rm {LSR}}=6.2$ km s$^{-1}$).\label{fig:f1}}
\end{figure}

We conducted data reduction using the Java Newstar\footnote{http://www.nro.nao.ac.jp/~jnewstar/html/}, which is software for data reduction and analyses of the Nobeyama data.
We fitted the spectra with a Gaussian profile and obtained spectral line parameters as summarized in Table \ref{tab:tab1}.
Figure \ref{fig:f1} shows the spectra of the three $^{13}$C isotopologues and the normal species of HC$_{3}$N in L1521B.
The spectra of the three $^{13}$C isotopologues were taken with the signal-to-noise ratios between 8.1 and 12.0.
The on source integration time is approximately 6 hours.
The $V_{\rm {LSR}}$ values of each line are well consistent with each other, and agree with the systemic velocity (6.5 km s$^{-1}$) within their errors.
The line widths ($\Delta v$) are also in good agreement with each other and previous results \citep{1992ApJ...392...551S, 2004ApJ...617..399H}.
The ratios of the integrated intensity ($\int T^{\ast}_{\mathrm A}dv$ K km s$^{-1}$) among the three $^{13}$C isotopologues are derived to be 0.95 ($\pm 0.13$) : 1.00 : 1.5 ($\pm 0.2$) ($1 \sigma$) for [H$^{13}$CCCN]:[HC$^{13}$CCN]:[HCC$^{13}$CN] in L1521B. 
The fractionation pattern is consistent with that in TMC-1 CP \citep{1998A&A...329.1156T}.

\begin{figure}
\figurenum{2}
\plotone{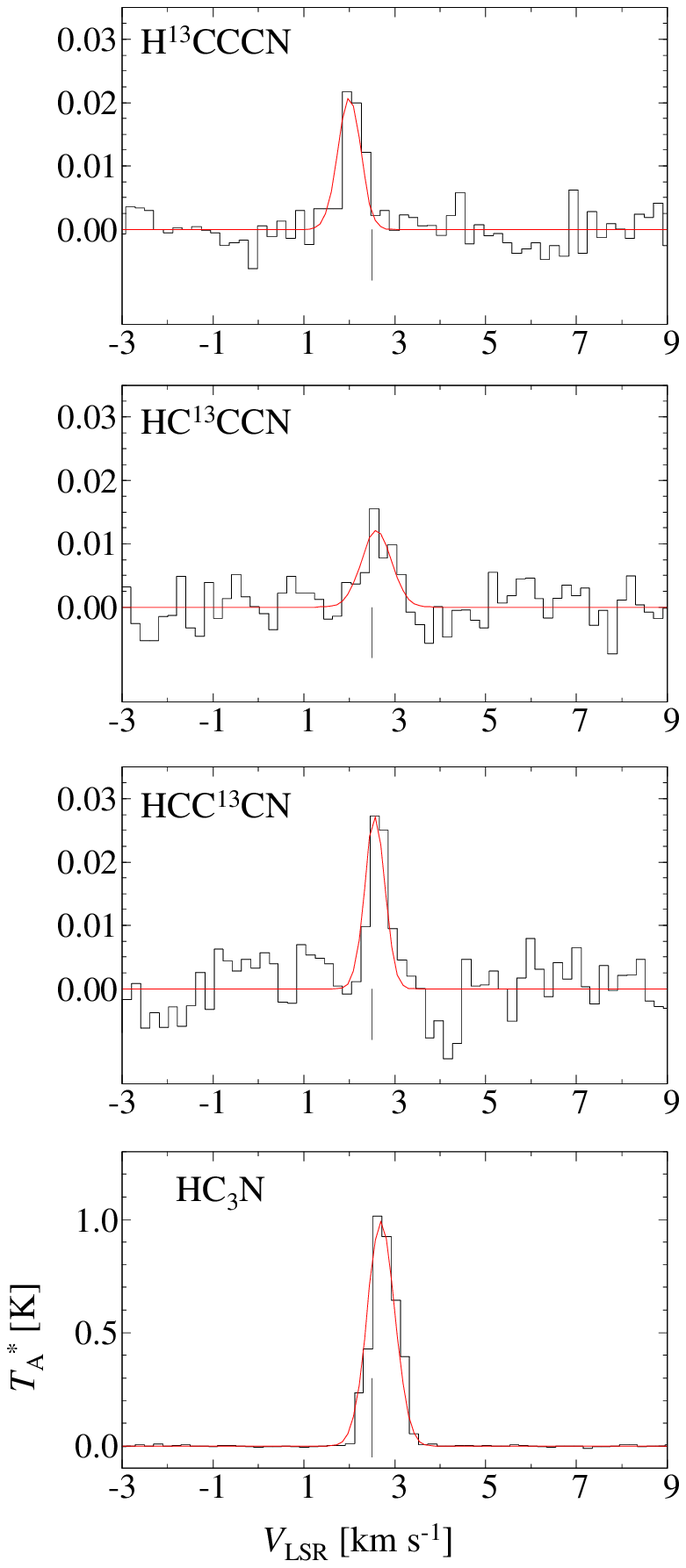}
\caption{Spectra of the normal species and the three $^{13}$C isotopologues of HC$_{3}$N in L134N. The vertical lines show the systemic velocity ($V_{\rm {LSR}}=2.5$ km s$^{-1}$).\label{fig:f2}}
\end{figure}

The spectra of the three $^{13}$C isotopologues in L134N were taken with the signal-to-noise ratios of 4.0$-$9.1, as shown in Figure \ref{fig:f2}.
The on source integration time is 38 hours and 18 minutes.
The $V_{\rm {LSR}}$ values are in good agreement with one another and the systemic velocity (2.5 km s$^{-1}$), while H$^{13}$CCCN shows a slightly smaller value but it is consistent within errors.
The line widths are consistent with each other and a previous result \citep{1992ApJ...392...551S}.
The ratios of the integrated intensity are found to be [H$^{13}$CCCN]:[HC$^{13}$CCN]:[HCC$^{13}$CN] = 1.5 ($\pm 0.2$) : 1.0 : 2.0 ($\pm 0.4$) ($1 \sigma$) in L134N.
This fractionation pattern is different from those in L1521B and TMC-1 CP \citep{1998A&A...329.1156T}.

%%%%%%%%%%%%%%%%%%%%%%%%%%%%%%%%%%%%%%%%%%%%%%%%%%%%%%%%%%%%%%%%%%%%%%%%%%%%%%%%%%%%%%%%%%%%%

%%%%%%%%%%%%%%%%%%%%%%%%%%%%%%%%%%%%%%%%%%%%%%%%%%%%%%%%%%%%%%%%%%%%%%%%%%%%%%%%%%%%%%%%%%%%%

%%%%%%%%%%%%%%%%%%%%%%%%%%%%%%%%%%%%%%%%%%%%%%%%%%%%%%%%%%%%%%%%%
\floattable
\begin{deluxetable}{lcccccccccccc}
\tabletypesize{\scriptsize}
\tablecaption{Spectral line parameters of HC$_{3}$N and its three $^{13}$C isotopologues in L1521B and L134N\label{tab:tab1}}
\tablewidth{0pt}
\tablehead{
\colhead{} & \colhead{} & \multicolumn{5}{c}{L1521B} & \colhead{} & \multicolumn{5}{c}{L134N}
\\
\cline{3-7}\cline{9-13}
\colhead{Species} & \colhead{Frequency\tablenotemark{a}} & \colhead{{\it T}$^{\ast}_{\mathrm A}$} & \colhead{$\Delta v$} & \colhead{$\int T^{\ast}_{\mathrm A}dv$} & \colhead{{\it V}$_{\mathrm {LSR}}$\tablenotemark{b}} & \colhead{rms\tablenotemark{c}} &  \colhead{} & \colhead{{\it T}$^{\ast}_{\mathrm A}$} & \colhead{$\Delta v$} & \colhead{$\int T^{\ast}_{\mathrm A}dv$} & \colhead{{\it V}$_{\mathrm {LSR}}$\tablenotemark{d}} & \colhead{rms\tablenotemark{c}} 
\\
\colhead{} & \colhead{(GHz)} & \colhead{(mK)} & \colhead{(km s$^{-1}$)} & \colhead{(K km s$^{-1}$)} & \colhead{(km s$^{-1}$)} & \colhead{(mK)} &  \colhead{} & \colhead{(mK)} & \colhead{(km s$^{-1}$)} & \colhead{(K km s$^{-1}$)} & \colhead{(km s$^{-1}$)} & \colhead{(mK)}
}
\startdata
HC$_{3}$N & 45.49031 & 2456 (82) & 0.73 (3) & 1.89 (10) & 6.3 & 7.9 & & 990 (26) & 0.71 (2) & 0.75 (3) & 2.7 & 3.4 \\
H$^{13}$CCCN & 44.08416 & 68 (7) & 0.67 (6) & 0.049 (7) & 6.3 & 6.5 & & 21 (2) & 0.67 (7) & 0.015(2) & 2.0 & 2.3 \\
HC$^{13}$CCN & 45.29733 & 61 (6) & 0.78 (8) & 0.051 (7) & 6.3 & 7.5 & & 12 (2) & 0.78 (17) & 0.010 (3) & 2.6 & 3.0 \\
HCC$^{13}$CN & 45.30171 & 102 (8) & 0.71 (5) & 0.077 (8) & 6.3 & 8.5 & & 27 (4) & 0.71 (9) & 0.020 (4) & 2.6 & 4.0 \\
\enddata
\tablecomments{Numbers in parentheses represent the standard deviation in the Gaussian fit, expressed in units of the last significant digits.}
\tablenotetext{a}{Taken from the Cologne Database for Molecular Spectroscopy (CDMS) \citep{2005JMoSt...742...215M}.}
\tablenotetext{b}{The errors are approximately 0.2 km s$^{-1}$, which correspond to the velocity resolution.}
\tablenotetext{c}{The rms noises in emission-free regions.}
\tablenotetext{d}{The errors are 0.5 km s$^{-2}$ at most due to smoothing in the velocity direction.}
\end{deluxetable}
%%%%%%%%%%%%%%%%%%%%%%%%%%%%%%%%%%%%%%%%%%%%%%%%%%%%%%%%%%%%%%%%%

\subsection{Analysis}

We derived the column densities of the normal species and the three $^{13}$C isotopologues of HC$_{3}$N assuming the local thermodynamic equilibrium using the following formulae \citep{2016ApJ...817...147T}:

\begin{equation} \label{tau}
\tau = - {\mathrm {ln}} \left[1- \frac{T^{\ast}_{\rm A} }{f\eta_{\rm B} \left\{J(T_{\rm {ex}}) - J(T_{\rm {bg}}) \right\}} \right],  
\end{equation}
where
\begin{equation} \label{tem}
J(T) = \frac{h\nu}{k}\Bigl\{\exp\Bigl(\frac{h\nu}{kT}\Bigr) -1\Bigr\} ^{-1},
\end{equation}  
and
\begin{equation} \label{col}
N = \tau \frac{3h\Delta v}{8\pi ^3}\sqrt{\frac{\pi}{4\mathrm {ln}2}}Q\frac{1}{\mu ^2}\frac{1}{J_{\rm {lower}}+1}\exp\Bigl(\frac{E_{\rm {lower}}}{kT_{\rm {ex}}}\Bigr)\Bigl\{1-\exp\Bigl(-\frac{h\nu }{kT_{\rm {ex}}}\Bigr)\Bigr\} ^{-1}.
\end{equation} 
In Equation (\ref{tau}), $\tau$ denotes the optical depth, $T^{\ast}_{\rm A}$ the antenna peak temperature (Table \ref{tab:tab1}), $f$ the beam filling factor, and $\eta_{\rm {B}}$ the main beam efficiency, respectively.
We used 1 for the beam filling factor in both L1521B and L134N, because the emission region sizes in L1521B \citep{2004ApJ...617..399H} and L134N \citep{2000ApJ...542..870D} are larger than the beam size of the Z45 receiver (37\arcsec, Section \ref{sec:obs}).
The main beam efficiency was 0.71 (Section \ref{sec:obs}).
$T_{\rm{ex}}$ and $T_{\rm {bg}}$ are the excitation temperature and the cosmic microwave background temperature ($\simeq 2.73$ K).
We assumed that the excitation temperature is 6.5 K \citep{1992ApJ...392...551S} for the normal species and the $^{13}$C isotopologues.
$J$($T$) in Equation (\ref{tem}) is the Planck function.
In Equation (\ref{col}), {\it N} is the column density,  $\Delta v$ is the line width (FWHM, Table \ref{tab:tab1}), $Q$ is the partition function, $\mu$ is the permanent electric dipole moment of HC$_{3}$N, and $E_{\rm {lower}}$ is the energy of the lower rotational energy level. 
We used 3.73172 D for $\mu$ of HC$_{3}$N \citep{1985JChPh...82...1702D}.

The derived column densities are summarized in Table \ref{tab:tab2}.
The optical depths are derived to be $3.01 \pm 0.15$, $0.0267 \pm 0.004$, $0.0241 \pm 0.003$, and $0.0403 \pm 0.004$ for HC$_{3}$N, H$^{13}$CCCN, HC$^{13}$CCN, and HCC$^{13}$CN, respectively, in L1521B.
The column density of the normal species derived here (($5.5 \pm 0.3$)$\times 10^{13}$ cm$^{-2}$) is slightly larger than that derived by \citet{1992ApJ...392...551S} ($4.1 \times 10^{13}$ cm$^{-2}$) by a factor of 1.3.
The difference can be explained by the uncertainty in the main beam efficiency.
If we used 0.75 for the main beam efficiency, which is the upper limit reported by the Nobeyama Radio Observatory\footnote{http://www.nro.nao.ac.jp/~nro45mrt/html/prop/status/Status\_R16.html}, the column density is derived to be ($4.2 \pm 0.2$)$\times 10^{13}$ cm$^{-2}$, and it is well consistent with that derived by \citet{1992ApJ...392...551S}.
The abundance ratios among the three $^{13}$C isotopologues are found to be 0.98 ($\pm 0.14$) : 1.00 : 1.52 ($\pm 0.16$) ($1 \sigma$) for [H$^{13}$CCCN]:[HC$^{13}$CCN]:[HCC$^{13}$CN].
 
In L134N, the optical depths are calculated to be $0.48 \pm 0.02$, $0.0081 \pm 0.0013$, $0.0047 \pm 0.0013$, and $0.011 \pm 0.002$ for HC$_{3}$N, H$^{13}$CCCN, HC$^{13}$CCN, and HCC$^{13}$CN, respectively.
There is no available literature deriving the column density of HC$_{3}$N by observations at the same position, and we do not discuss comparison between our results and others.
The abundance ratios among the three $^{13}$C isotopologues are derived to be [H$^{13}$CCCN]:[HC$^{13}$CCN]:[HCC$^{13}$CN] =  1.5 ($\pm 0.2$) : 1.0 : 2.1 ($\pm 0.4$) ($1 \sigma$).

%%%%%%%%%%%%%%%%%%%%%%%%%%%%%%%%%%%%%%%%%%%%%%%%%%%%%%%%%%%%%%%%%
\floattable
\begin{deluxetable}{cccccc}
\tabletypesize{\scriptsize}
\tablecaption{Column densities of HC$_{3}$N and its three $^{13}$C isotopologues and the $^{12}$C/$^{13}$C ratios in L1521B and L134N\label{tab:tab2}}
\tablewidth{0pt}
\tablehead{
\colhead{} & \multicolumn{2}{c}{L1521B} & \colhead{} & \multicolumn{2}{c}{L134N} \\
\cline{2-3}\cline{5-6}
\colhead{Species} & \colhead{Column density} & \colhead{$^{12}$C/$^{13}$C} & \colhead{} & \colhead{Column density} & \colhead{$^{12}$C/$^{13}$C} \\
\colhead{} & \colhead{($\times 10^{11}$ cm$^{-2}$)} & \colhead{ratio} & \colhead{} & \colhead{($\times 10^{11}$ cm$^{-2}$)} & \colhead{ratio}
}
\startdata
HC$_{3}$N & ($5.5 \pm 0.3$)$\times 100$ & $-$ & & ($8.6 \pm 0.3$)$\times 10$ & $-$ \\
H$^{13}$CCCN & $4.7 \pm 0.6$ & $117 \pm 16$ & & $1.4 \pm 0.2$ & $61 \pm 9$ \\
HC$^{13}$CCN & $4.7 \pm 0.7$ & $115 \pm 16$ & & $0.9 \pm 0.2$ & $94 \pm 26$ \\
HCC$^{13}$CN & $7.2 \pm 0.8$ & $76 \pm 6$ & & $1.9 \pm 0.3$ & $46 \pm 9$ \\ 
\enddata
\tablecomments{Errors represent the standard deviation.}
\end{deluxetable}
%%%%%%%%%%%%%%%%%%%%%%%%%%%%%%%%%%%%%%%%%%%%%%%%%%%%%%%%%%%%%%%%%

\section{Discussion} \label{sec:dis}

\subsection{Main formation pathways of HC$_{3}$N in L1521B and L134N} \label{sec:d1}

The differences in abundance among the three $^{13}$C isotopologues of HC$_{3}$N, namely the $^{13}$C isotopic fractionation, cannot be brought by the isotope exchange reactions, as discussed by \citet{1998A&A...329.1156T} in detail.
\citet{2006JChPh.124d4307L} showed that the reaction between HC$_{3}$N and carbon atom is efficient at the temperature as low as 10 K.
They suggested that the reaction of ``$^{13}$C + HC$_{3}$N" may form various $^{13}$C isotopologues of HC$_{3}$N.
However, it is not still clear whether the reaction contributes to the $^{13}$C isotopic fractionation of HC$_{3}$N, because their results did not clearly show which $^{13}$C isotopologue is selectively formed. 
We then will not consider the effects of the reaction on the $^{13}$C isotopic fractionation of HC$_{3}$N in the following discussion.
Hence, the observed $^{13}$C isotopic fractionation of HC$_{3}$N should occur during its formation processes and reflect its main formation pathways.
We investigate the main formation pathways of HC$_{3}$N in each starless core in this section based on the observed $^{13}$C isotopic fractionation.

\begin{figure}
\figurenum{3}
\plotone{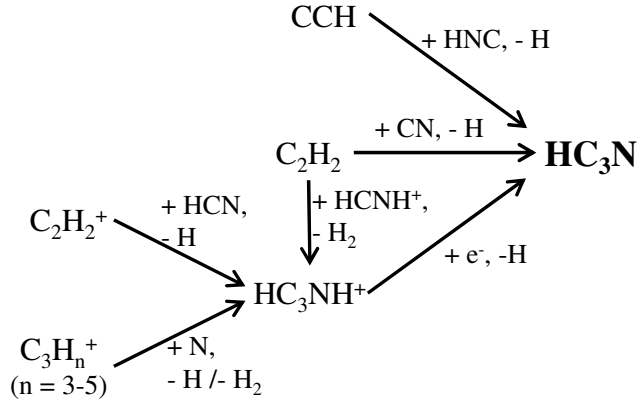}
\caption{Possible main formation pathways leading to HC$_{3}$N.\label{fig:f3}}
\end{figure}

We find out that there are three possible main routes leading to HC$_{3}$N \citep{2016ApJ...830...106T}, using the UMIST Database for Astrochemistry 2012 \citep{2013A&A...550A..36M} as follows.\\
\noindent Pathway 1: the neutral-neutral reaction between C$_{2}$H$_{2}$ and CN\\
\noindent Pathway 2: the neutral-neutral reaction between CCH and HNC\\
\noindent Pathway 3: the electron recombination reaction of HC$_{3}$NH$^{+}$\\
Figure \ref{fig:f3} shows the reaction schemes of the formation pathways leading to HC$_{3}$N.
We add the reaction of ``C$_{2}$H$_{2}$ + HCNH$^{+}$ $\rightarrow $ HC$_{3}$NH$^{+}$ + H$_{2}$" \citep{1979ApJ...233..102M}, besides the UMIST 2012 database.

As discussed by \citet{2016ApJ...830...106T} in detail, the predicted $^{13}$C isotopic fractionation patterns brought from each possible main formation mechanism are as follows;\\
\noindent Pathway 1: [H$^{13}$CCCN]:[HC$^{13}$CCN]:[HCC$^{13}$CN] $= 1 : 1: x$ ($x$ is an arbitrary value) \\
\noindent Pathway 2: [H$^{13}$CCCN]:[HC$^{13}$CCN]:[HCC$^{13}$CN] = $x : 1: y$ ($x$ and $y$ are arbitrary values) \\
\noindent Pathway 3: [H$^{13}$CCCN]:[HC$^{13}$CCN]:[HCC$^{13}$CN] $\approx 1 : 1: 1$ \\
%If the main formation pathway of HC$_{3}$N is Pathway 1 (C$_{2}$H$_{2}$ + CN), the abundance ratios of the three $^{13}$C isotopologues should be [H$^{13}$CCCN]:[HC$^{13}$CCN]:[HCC$^{13}$CN] $= 1 : 1: x$, where $x$ is an arbitrary value.
%These expected ratios can be explained by the two following reasons:
%\begin{enumerate}
%\item C$_{2}$H$_{2}$ molecule has two equivalent carbon atoms, and
%\item the triple bond between carbon and nitrogen atoms in CN molecule is preserved during the reaction process \citep{1997ApJ...489..113F}.
%\end{enumerate}
%
%If the main formation pathway is Pathway 2 (CCH + HNC), the abundance ratios of the three $^{13}$C isotopologues of HC$_{3}$N should be [H$^{13}$CCCN]:[HC$^{13}$CCN]:[HCC$^{13}$CN] = $x : 1: y$, where $x$ and $y$ are arbitrary values.
%In this reaction, a carbon atom in HNC connects to a carbon atom with an unpaired electron in CCH  \citep{1997ApJ...489..113F}.
%\citet{2010ApJ...512...A31S} showed that C$^{13}$CH is more abundant than $^{13}$CCH by a factor of 1.6 in TMC-1 CP and L1527.

Regarding Pathway 2, \citet{2011ApJ...731...38F} found that the difference in abundance between C$^{13}$CH and $^{13}$CCH arises mostly due to the exchange reaction, $^{13}$CCH + H $\rightleftharpoons$ C$^{13}$CH + H + $\Delta E$ (8.1 K), rather than during its formation process.
Therefore, the $^{13}$C isotopic fractionation in CCH could occur independently from its formation pathways.
The $^{13}$C isotopic fractionation in CCH would be preserved during the reaction of Pathway 2 \citep{1997ApJ...489..113F}, and $x$ is expected to be larger than 1.

In case of Pathway 3, the $^{13}$C isotopic fractionation in HC$_{3}$NH$^{+}$ will be averaged by several formation processes of the ion, as shown in Figure \ref{fig:f3}.
In addition, there is no reason that $^{13}$C concentrates in a particular carbon atom in HC$_{3}$NH$^{+}$ after the ion is formed.
%If Pathway 3 (HC$_{3}$NH$^{+}$ + e$^{-}$) is the main formation pathway of HC$_{3}$N, there is possibly no significant difference in abundance among all of the $^{13}$C isotopologues of HC$_{3}$N, that is [H$^{13}$CCCN]:[HC$^{13}$CCN]:[HCC$^{13}$CN] $\approx 1 : 1: 1$.
%HC$_{3}$NH$^{+}$ is formed by several processes as shown in Figure \ref{fig:f3}.
%In that case, the $^{13}$C isotopic fractionation in HC$_{3}$NH$^{+}$ will be averaged \citep{2016ApJ...817...147T}.
%Scrambling may occur during the ion-molecule reactions, which leads to averaging the $^{13}$C isotopic fractionation \citep{2016ApJ...830...106T}.
%In addition, there is no reason that $^{13}$C concentrates in a particular carbon atom in HC$_{3}$NH$^{+}$ after the ion is formed.
%In fact, \citet{2016ApJ...817...147T} proposed that the main formation mechanism of HC$_{5}$N in TMC-1 CP is the ion-molecule reactions between hydrocarbon ions and nitrogen atoms followed by the electron recombination reactions, based on the results of no clear $^{13}$C isotopic fractionation among the five $^{13}$C isotopologues.

From comparisons between the observational results in L1521B and L134N with the above expected fractionation patterns, we propose possible main formation pathways of HC$_{3}$N in each low-mass starless core as follows.
\begin{enumerate}
\item L1521B : the main formation pathway of HC$_{3}$N is the reaction of C$_{2}$H$_{2}$ + CN (Pathway 1).
\item L134N : the main formation pathway of HC$_{3}$N is the reaction of CCH + HNC (Pathway 2).
\end{enumerate}

According to Figure 6 in \citet{2011ApJ...731...38F}, the both CCH/$^{13}$CCH and CCH/C$^{13}$CH ratios steeply increase and decrease at the early stage around 10$^{3}$ yr, and the ratios are almost constant by $\sim10^{5}$ yr.
The derived $^{12}$C/$^{13}$C ratios of HC$_{3}$N in L134N summarized in Table \ref{tab:tab2} agree with their model at just before 10$^{5}$ yr (CCH/$^{13}$CCH $\sim 100$ and CCH/C$^{13}$CH $\sim 60$).
The C$^{13}$CH/$^{13}$CCH ratio \citep[Figure 3 (b) in][]{2011ApJ...731...38F} is approximately 1.6 between 10$^{4}$ and 10$^{5}$ yr and starts to increase just before 10$^{5}$ yr.
The H$^{13}$CCCN/HC$^{13}$CCN ratio is derived to be $1.5 \pm 0.2$. 
If the main formation pathway of HC$_{3}$N is the neutral-neutral reaction of CCH + HNC, the expected C$^{13}$CH/$^{13}$CCH ratio is $1.5 \pm 0.2$, which is consistent with 1.6 calculated by \citet{2011ApJ...731...38F}.

%%%%%%%%%%%%%%%%%%%%%%%%%%%%%%%%%%%%%%%%%
%%%%%%%%%%%%%%%%%%%%%%%%%%%%%%%%%%%%%%%%%

\subsection{What a factor contributes the differences in main formation pathway between L134N and L1521B/TMC-1 CP?} \label{sec:d3}

\citet{2016ApJ...830...106T} compared the $^{13}$C isotopic fractionation of HC$_{3}$N in various sources from low-mass prestellar to high-mass star-forming cores.
They found that the fractionation patterns in all of the four sources are the same, and suggested that the primary formation pathway of HC$_{3}$N may be common in various physical conditions.
Chemical reactions significantly depend on the physical conditions such as temperature and density.
It is then naturally expected that the $^{13}$C isotopic fractionation patterns in the similar physical conditions are the same.
The temperature and density are almost the same among the three clouds, L134N, L1521B, and TMC-1 CP ($T_{\rm kin} \sim 10$ K, $n \sim 10^{4} - 10^{5}$ cm$^{-3}$).
Nevertheless, the suggested main formation pathway in L134N is different from those in L1521B and TMC-1 CP (Section \ref{sec:d1}).
It is unclear what a factor brings the difference between L134N and L1521B/TMC-1 CP, and we discuss the matter in this section.

The age of clouds affects the chemical composition in starless cores, namely the chemical evolution.
\citet{2004ApJ...617..399H} suggested that L1521B is in a very early stage of physical and chemical evolution.
On the other hand, \citet{2000ApJ...542..870D} suggested that L134N is chemically evolved than TMC-1 CP, because carbon-chain species in L134N are less than in TMC-1 CP.
In addition, \citet{2009ApJ...699..585H} found that the NH$_{3}$/CCS abundance ratio in L134N is high (444) suggestive of a chemically evolved core.
Using the results of \citet{1992ApJ...392...551S}, we calculated the NH$_{3}$/CCS abundance ratios in L1521B and TMC-1 CP to be 1.7 and 2.9, respectively, suggesting that the both starless cores are chemically young.
In summary, the chemical age of L1521B is comparable with that of TMC-1 CP, and L134N is more evolved than the two cores.

We found the following three changes depending on the age of clouds with regard to our observational results and formation pathways of HC$_{3}$N.

\begin{enumerate}
\item the $^{12}$C/$^{13}$C ratios decrease,
\item the CCH abundance decreases, and
\item the CN/HNC ratio decreases.
\end{enumerate}

The first one, decreasing in the $^{12}$C/$^{13}$C ratios, does not affect the formation pathways of HC$_{3}$N, but it is a key for cloud evolution.
\citet{2011ApJ...731...38F} demonstrated their chemical model calculations taking into consideration of the $^{13}$C isotopic fractionation by isotopomer-exchange reactions.
They showed the time dependences of the $^{13}$C isotopic fractionation and the $^{12}$C/$^{13}$C ratios.
In the case of $n_{\rm {H}_{2}} = 5 \times 10^{3}$ cm$^{-3}$, the $^{12}$C/$^{13}$C ratios of HC$_{3}$N were derived to be 123, 85, and 99 at 10$^{4}$, 10$^{5}$, and 10$^{6}$ yr, respectively.
The $^{12}$C/$^{13}$C ratios of HC$_{3}$N in L1521B is largely higher than those in L134N, as summarized in Table \ref{tab:tab2}.
The high $^{12}$C/$^{13}$C ratios of HC$_{3}$N observed in L1521B may imply the very young starless core as suggested by \citet{2004ApJ...617..399H}, while its low ratios observed in L134N seem to indicate the evolved starless core as suggested by \citet{2000ApJ...542..870D}.

The second and third ones are related to the reaction rates leading to HC$_{3}$N.
Both C$_{2}$H$_{2}$ and CCH are mainly produced by the electron recombination reactions of C$_{2}$H$_{3}^{+}$ \citep[e.g.][]{2013ChRv...113...8981S} and they have a chemically close relationship.
The CCH abundance decrease after 10$^{5}$ yr, while C$_{2}$H$_{2}$ does not decrease significantly\footnote{http://udfa.ajmarkwick.net/index.php}.
In that case, chemically evolved cores are poorer in CCH than in younger cores, and the reaction of CCH + HNC will be less effective.
Taking into consideration that L134N is a more evolved core, the low abundance of CCH in evolved cores cannot explain the observational results and our suggestions.
Therefore, it is unlikely that the difference between CCH and C$_{2}$H$_{2}$ brings the difference between L134N and L1521B/TMC-1 CP.

\begin{figure}
\figurenum{4}
\plotone{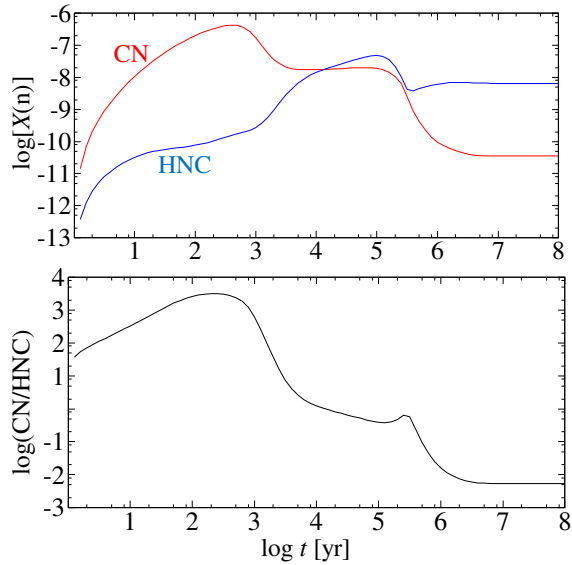}
\caption{Time dependence of abundances of CN and HNC (upper). Time dependence of the CN/HNC ratio (lower).\label{fig:f4}}
\end{figure}

The abundances of CN and HNC increase and reach at the peak around 10$^{3}$ yr and 10$^{5}$ yr, respectively.
Figure \ref{fig:f4} shows the time dependence of the abundances of CN and HNC.
We run chemical network model calculation simply using the dataset of dark clouds models provided by the UMIST Database for Astrochemistry 2012\footnote{http://udfa.ajmarkwick.net/index.php?mode=downloads} \citep{2013A&A...550A..36M}, assuming that temperature, density, and visual extinction are 10 K, $2 \times 10^{4}$ cm$^{-3}$, and 10 magnitude, respectively.
We also derive the CN/HNC ratio as shown in Figure \ref{fig:f4}.
The CN/HNC ratio starts to decrease before 10$^{3}$ yr, and the ratios at 10$^{4}$ yr and 10$^{5}$ yr are lower than the peak value ($t =2 \times 10^{2}$ yr) by approximately three orders of magnitude and four orders of magnitude, respectively.
L134N is chemically evolved, and then CN may be already depleted.
In fact, the CN/HNC ratios are calculated to be 0.028 (HNC/CN = 35.6) in TMC-1 CP from the results of \citet{1997ApJ...486..862P} and 0.018 (HNC/CN = 54.2) in L134N from the results of \citet{2000ApJ...542..870D}.
In addition, L134N is considered to be rich in oxygen \citep[e.g.][]{2014MNRAS.437..930L}.
In that case, CN seems to be further destroyed by the reaction with oxygen atom to produce CO \citep{2014MNRAS.437..930L}.
To summarize, there is a possibility that the different main formation pathways of HC$_{3}$N in L134N and L1521B/TMC-1 CP arise from the different CN/HNC ratios.
\\
%%%%%%%%%%%%%%%%%%%%%%%%%%%%%%%%%%%%%%%%%
%%%%%%%%%%%%%%%%%%%%%%%%%%%%%%%%%%%%%%%%% 
\\
\\
\section{Conclusions} \label{sec:con}

We carried out simultaneous observations of the normal species and the three $^{13}$C isotopologues of HC$_{3}$N at the 45 GHz band toward L1521B and L134N with the Nobeyama 45 m radio telescope.
In L1521B, the abundance ratios among the three $^{13}$C isotopologues are derived to be [H$^{13}$CCCN]:[HC$^{13}$CCN]:[HCC$^{13}$CN] = 0.98 ($\pm 0.14$) : 1.00 : 1.52 ($\pm 0.16$) ($1 \sigma$).
From the fractionation pattern, we propose that the main formation pathway of HC$_{3}$N is the neutral-neutral reaction between C$_{2}$H$_{2}$ and CN, which is the same one as in TMC-1 CP.
On the other hand, the abundance ratios in L134N are [H$^{13}$CCCN]:[HC$^{13}$CCN]:[HCC$^{13}$CN]= 1.5 ($\pm 0.2$) : 1.0 : 2.1 ($\pm 0.4$) ($1 \sigma$).
This is a different fractionation pattern.
From comparison of the expected fractionation pattern of the possible efficient formation pathways of HC$_{3}$N in dark clouds, we found that the fractionation pattern in L134N agrees with the reaction between CCH and HNC.
Although the physical conditions are similar in these dark clouds, the possible main formation pathway in L134N is different from those suggested in L1521B and TMC-1 CP.
This is the first observational studies to show that HC$_{3}$N is formed differently among the dark clouds with the similar physical conditions.
The low CN/HNC ratio in L134N may cause the differences between L134N and L1521B/TMC-1 CP.

\acknowledgments
We would like to express our thanks to the staff of the Nobeyama Radio Observatory.
In particular, we deeply appreciate a team mainly led by Dr. Yusuke Miyamoto, Dr. Tetsuhiro Minamidani, Dr. Mitsuhiro Matsuo, Mr. Tomio Kanzawa, and Mr. Jun Maekawa for evaluating pointing errors without the master collimator driving system and quick restarting operation.
The Nobeyama Radio Observatory is a branch of the National Astronomical Observatory of Japan, National Institutes of Natural Sciences.
The Z45 receiver is supported in part by a Granting-Aid for Science Research of Japan (24244017).


\vspace{5mm}
\facilities{Nobeyama 45 m radio telescope}


\software{Java Newstar}

\end{document}